\documentclass[a4paper]{jpconf}
\usepackage{graphicx,iopams}
\usepackage{amsmath,bm}
\begin{document}
\title{Ferromagnetism and orbital order in the two-orbital Hubbard model}

\author{Katsunori Kubo}

\address{Max Planck Institute for Chemical Physics of Solids,
  01187 Dresden, Germany}
\address{Advanced Science Research Center, Japan Atomic Energy Agency,
  Tokai, Ibaraki 319-1195, Japan}

\vspace{5.5mm}

\begin{abstract}
We investigate spin and orbital states of the two-orbital Hubbard model
on a square lattice by using a variational Monte Carlo method
at quarter-filling, i.e., the electron number per site is one.
As a variational wave function,
we consider a Gutzwiller projected wave function of a mean-field type
wave function for a staggered spin and/or orbital ordered state.
Then, we evaluate expectation value of energy for the variational wave functions
by using the Monte Carlo method and determine the ground state.
In the strong Coulomb interaction region,
the ground state is the perfect ferromagnetic state with antiferro-orbital (AF-orbital) order.
By decreasing the interaction,
we find that the disordered state becomes the ground state.
Although we have also considered the paramagnetic state with AF-orbital order,
i.e., purely orbital ordered state,
and partial ferromagnetic states with and without AF-orbital order,
they do not become the ground state.
\end{abstract}

\section{Introduction}
It has been recognized that orbital degree of freedom
is important for magnetism in some materials,
such as colossal magneto-resistance and complex ordered phases
of manganites~\cite{Imada,Hotta},
and exotic magnetism in $f$-electron systems~\cite{Hotta,Santini}.
The orbital degree of freedom plays an important role
in particular for realization of ferromagnetism.
In two orbital models at quarter-filling, i.e., the electron number per site is one,
a perfect ferromagnetic state realizes with antiferro-orbital (AF-orbital) order in the strong coupling limit~\cite{Kugel}.
On the other hand, in the intermediate coupling regime,
it is not obvious whether the inclusion of the orbital degree of freedom
is enough for the stabilization of ferromagnetism.
In addition, it is interesting whether the perfect ferromagnetic state
is continuously connected to the paramagentic state
through partial ferromagnetic states or
the ground state changes abruptly from the perfect ferromagnetic state
to the paramagnetic state at a certain interaction strength.

In this study,
in order to unveil the ground state property
in a system with the orbital degree of freedom
in weak coupling to strong coupling cases,
we study spin and orbital states of the two-orbital Hubbard model
by using a variational Monte Carlo method at quarter-filling.
A similar model has been studied by the variational Monte Carlo method
for states without orbital order, and it has been found that
staggered orbital correlation enhances in a ferromagnetic state
in a strong coupling region~\cite{Kobayashi1}.
In the strong Coulomb interaction limit,
the ground state should be the ferromagnetic state with AF-orbital order,
and in the weak coupling case, the ground state is the disordered state.
In this study, in addition to these two states, we consider
a purely AF-orbital ordered state without magnetic order,
a purely magnetic ordered state without orbital order,
and partial ferromagnetic state with and without AF-orbital order.

\section{Model and Method}
The two-orbital Hubbard model is given by the following Hamiltonian:
\begin{equation}
  \begin{split}
  H=&\sum_{\bm{k} \tau \sigma} \epsilon_{\bm{k}}c^{\dagger}_{\bm{k} \tau \sigma}c_{\bm{k} \tau \sigma}
  +U \sum_{i, \tau} n_{i \tau \uparrow} n_{i \tau \downarrow}
  +U^{\prime} \sum_{i} n_{i 1} n_{i 2}\\
  &+J\sum_{i,\sigma,\sigma^{\prime}}
    c^{\dagger}_{i 1 \sigma} c^{\dagger}_{i 2 \sigma^{\prime}}
    c_{i 1 \sigma^{\prime}} c_{i 2 \sigma}
  +J^{\prime}\sum_{i,\tau \ne \tau^{\prime}}
    c^{\dagger}_{i \tau \uparrow} c^{\dagger}_{i \tau \downarrow}
    c_{i \tau^{\prime} \downarrow} c_{i \tau^{\prime} \uparrow},
    \end{split}
\end{equation}
where $c_{i\tau\sigma}$ is the annihilation operator of
the electron at site $i$ with orbital $\tau$ ($=1$ or $2$)
and spin $\sigma$ ($=\uparrow$ or $\downarrow$),
$c_{\bm{k}\tau\sigma}$ is the Fourier transform of it,
$n_{i \tau \sigma}=c^{\dagger}_{i \tau \sigma} c_{i \tau \sigma}$, and
$n_{i \tau}=\sum_{\sigma}n_{i \tau \sigma}$.
We consider the nearest neighbor hopping and the dispersion
is given by $\epsilon_{\bm{k}}=-2t(\cos k_x+\cos k_y)$.
We set the lattice constant unity.
The coupling constants $U$, $U^{\prime}$, $J$, and $J^{\prime}$
denote the intra-orbital Coulomb, inter-orbital Coulomb, exchange,
and pair-hopping interactions, respectively.

We evaluate expectation value of energy for variational wave functions
by using the Monte Carlo method.
We consider a Gutzwiller-projected wave function
of a mean-field type wave function
as a variational wave function~\cite{Kobayashi1,Bunemann,Kobayashi2}.
In this study,
we consider staggered order of spin and/or orbital states
for the mean-field type wave function.
The variational wave function is given by
\begin{equation}
  | \Psi \rangle = P_{\text{G}} | \Phi \rangle,
\end{equation}
where $P_{\text{G}}$ is the Gutzwiller projection operator defined in Ref~\cite{Kobayashi2}.
The mean-field type wave function $| \Phi \rangle$ is given by
\begin{equation}
  |\Phi \rangle = \prod_{\bm{k} m \tau \sigma}
  b^{(m) \dagger}_{\bm{k} \tau \sigma} |0 \rangle,
\end{equation}
where $| 0 \rangle$ is the vacuum.
The quasiparticles occupy $N_{\sigma}$ states for each spin $\sigma$
from the lowest quasiparticle energy state,
where $N_{\sigma}$ is the number of electrons with spin $\sigma$.
The number of parameters in the Gutzwiller projection
can be reduced from sixteen to ten by considering symmetry~\cite{Kobayashi2},
and we can further reduce the number of the parameters to seven
when we fix $N_{\sigma}$.
The energy of the quasiparticle in the ordered state is given by
\begin{equation}
  \lambda^{(m)}_{\bm{k} \tau \sigma}=m\sqrt{\Delta^2_{\tau \sigma}+\epsilon^2_{\bm{k}}}.
\end{equation}
The creation operators of quasiparticles are given by
\begin{align}
  b^{(-) \dagger}_{\bm{k} \tau \sigma}
  &=u_{\bm{k} \tau \sigma} c^{\dagger}_{\bm{k} \tau \sigma}
  +\text{sgn}(\Delta_{\tau \sigma})v_{\bm{k} \tau \sigma} c^{\dagger}_{\bm{k}+\bm{Q} \tau \sigma},\\
  b^{(+) \dagger}_{\bm{k} \tau \sigma}
  &=-\text{sgn}(\Delta_{\tau \sigma})v_{\bm{k} \tau \sigma} c^{\dagger}_{\bm{k} \tau \sigma}
  +u_{\bm{k} \tau \sigma} c^{\dagger}_{\bm{k}+\bm{Q} \tau \sigma},
\end{align}
where $\bm{Q}=(\pi,\pi)$ is the ordering vector considered in this study and
\begin{align}
  u_{\bm{k} \tau \sigma}=\left[ \left( 1-\epsilon_{\bm{k}}/\sqrt{\Delta^2_{\tau \sigma}+\epsilon^2_{\bm{k}}} \right) /2 \right]^{1/2},\\
  v_{\bm{k} \tau \sigma}=\left[ \left( 1+\epsilon_{\bm{k}}/\sqrt{\Delta^2_{\tau \sigma}+\epsilon^2_{\bm{k}}} \right) /2 \right]^{1/2}.
\end{align}
The quasiparticle gap in the ordered state is given by
\begin{equation}
  \Delta_{\tau \sigma}=\Delta_{\text{c}}+\Delta_{\text{s}}(\delta_{\sigma \uparrow}-\delta_{\sigma \downarrow})
  +\Delta_{\text{o}}(\delta_{\tau 1}-\delta_{\tau 2})+\Delta_{\text{so}}(\delta_{\sigma \uparrow}-\delta_{\sigma \downarrow})(\delta_{\tau 1}-\delta_{\tau 2}),
\end{equation}
where $\Delta_{\text{c}}$, $\Delta_{\text{s}}$, $\Delta_{\text{o}}$, and $\Delta_{\text{so}}$
denote the gap for charge, spin, orbital, and spin-orbital ordered states, respectively.
We evaluate the expectation value of energy for the variational wave function
by using the Monte Carlo method,
and optimize these gap parameters and the Gutzwiller parameters
to minimize energy.
We can also evaluate energy by fixing some parameters, for example,
we set all the gap parameters zero for the disordered phase.
In addition to states without magnetization,
we also consider states with finite magnetization
$m=(N_{\uparrow}-N_{\downarrow})/(N_{\uparrow}+N_{\downarrow})$.

\section{Result}
\begin{figure}[b]
\begin{center}
  \includegraphics[width=15cm]{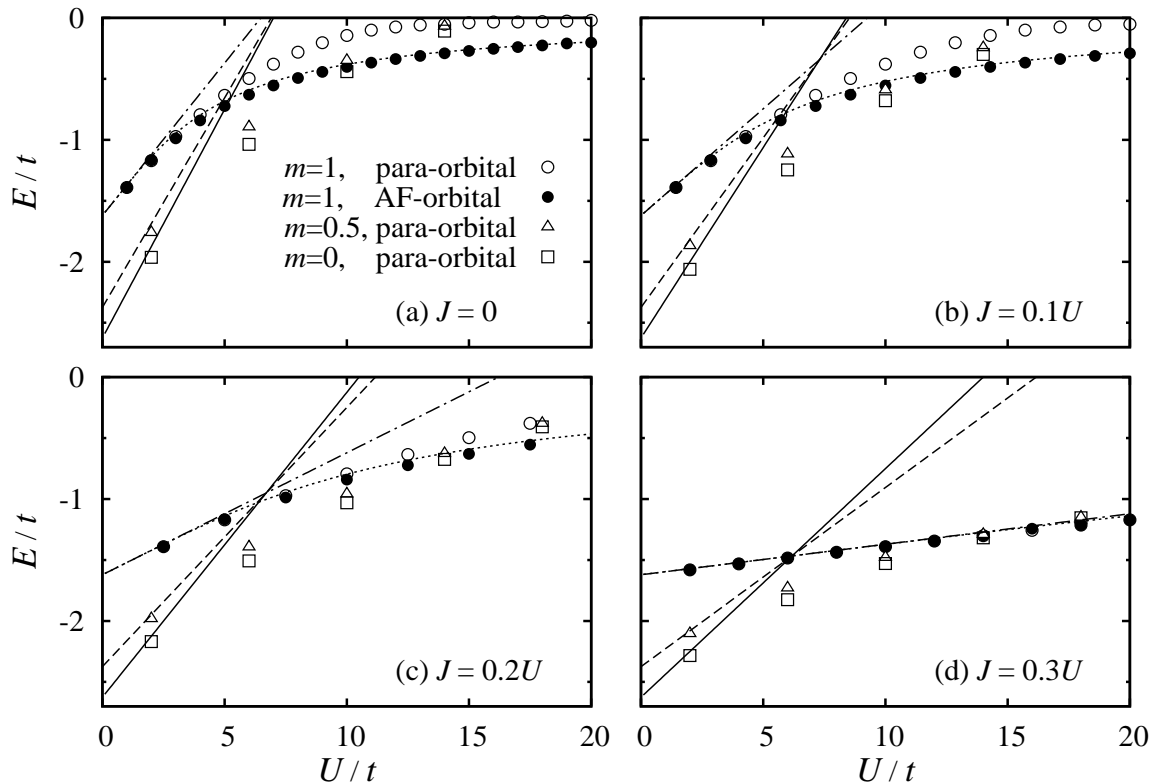}
\end{center}
\caption{\label{fig}
Energy as functions of the Coulomb interaction
obtained with several variational functions
(a) for $J=0$, (b) for $J=0.1U$, (c) for $J=0.2U$, and (d) for $J=0.3U$.
Dash-dotted line, dotted line, dashed line, and solid line
represent energy of Hartree-Fock wave functions for
$m=1$ para-orbital,
$m=1$ AF-orbital,
$m=0.5$ para-orbital, and
$m=0$ para-orbital,
respectively.}
\end{figure}
Figure~\ref{fig} shows energy $E$ per site as functions of $U$
of several states for $J=0$, $J=0.1U$, $J=0.2U$, and $J=0.3U$
on a $12 \times 12$ lattice.
We have used the relations $U=U^{\prime}+J+J^{\prime}$ and $J=J^{\prime}$.
Note that at $J=0$, the orbital space is equivalent to the spin space
and we can exchange them, for example,
the ferromagnetic AF-orbital state is equivalent to antiferro-spin ferro-orbital state.
At $J \ne 0$, the orbital space is not equivalent to the spin space,
and the ground state is uniquely determined except for trivial degeneracy,
e.g., the degeneracy due to the rotational symmetry in the spin space.
The lines are the Hartree-Fock energy, i.e., without the Gutzwiller projection.

We find that the paramagnetic ($m=0$) orbital-disordered (para-orbital)
state has the lowest energy for $U < U_{\text{c}}$,
where $U_{\text{c}} \simeq 10t$--$15t$.
For $U > U_{\text{c}}$, the perfect ferromagnetic ($m=1$) AF-orbital state
becomes the ground state.
When the Coulomb interaction becomes large, electrons tend to occupy the
same spin state to avoid the large energy loss by the Coulomb interaction
even though the kinetic energy becomes large.
In the perfect ferromagnetic state, the model reduces to the single-orbital
Hubbard model at half-filling with Coulomb interaction
$U_{\text{eff}}=U^{\prime}-J$
if we regard the spin of the single-orbital Hubbard model as orbital.
Thus, in the perfect ferromagnetic state, orbital-AF order occurs to reduce energy.
The obtained critical value $U_{\text{c}}$
from the disordered state to the ferromagnetic state is not so large.
If we do not consider the orbital order, $U_{\text{c}}$ becomes larger.
The critical value $U_{\text{c}}$ increases as $J$ is increased.
By increasing $J$, the effective interaction $U_{\text{eff}}$ between different orbitals
becomes weak, and a larger value of the Coulomb interaction is necessary for
the appearance of the ferromagnetic state.

Although we have also calculated energy for the purely AF-orbital state with $m=0$
and for partial ferromagnetic states, e.g., $m=0.5$ shown in figure~\ref{fig},
with and without AF-orbital order,
energy for them does not become the lowest.
This fact indicates that the ferromagnetism and the AF-orbital order stabilize each other
and occur simultaneously in the ground state at quarter-filling.
Note that the absence of a partial ferromagnetic state at quarter-filling
has also been found in one dimension~\cite{Sakamoto}
and in infinite dimensions~\cite{Momoi}.

\section{Summary}
We have studied ground state of the two-orbital Hubbard model on a square lattice
at quarter-filling within the variational wave functions.
For the variational wave functions, we have considered
Gutzwiller projected functions of spin and/or orbital ordered states.
We have found that in the strong coupling region,
the ground state is the perfect ferromagnetic state with AF-orbital order.
In the weak coupling region, the ground state is the disordered state.
We have also found that, at quarter-filling,
other states do not become the ground state
even in the intermediate coupling region.

\ack
This work is supported by
JSPS Postdoctoral Fellowships for Research Abroad.

\end{document}